\documentclass[runningheads]{llncs}
\usepackage{graphicx}
\usepackage{multirow}

\usepackage[caption=false]{subfig}
\usepackage{rotating}

\begin{document}
%
\title{Bioelectric Registration of Electromagnetic Tracking and Preoperative Volume Data}
\titlerunning{Bioelectric Registration of EM Tracking and Preoperative Volume Data}
%

\author{
	Ardit Ramadani\inst{1, 2, 3, *} \and
	Heiko Maier \inst{1, 2, 3, *} \and
	Felix Bourier MD \inst{2} \and
	Christian Meierhofer MD \inst{2} \and
	Peter Ewert MD \inst{2} \and
	Heribert Schunkert MD \inst{2, 3} \and
	Nassir Navab\inst{1, 4, 5}
}

\authorrunning{A. Ramadani, H. Maier et al.}

	
\institute{
	Computer Aided Medical Procedures,\\ Technical University of Munich, Munich, Germany\\ 
	\email{ardit.ramadani@tum.de, heiko.maier@tum.de}\\
	\and German Heart Center Munich, Munich, Germany
	\and Deutsches Zentrum f\"ur Herz- und Kreislaufforschung (DZHK),\\Munich Heart Alliance, Germany
	\and Munich Institute of Robotics and Machine Intelligence,\\ Technical University of Munich, Munich, Germany
	\and Computer Aided Medical Procedures,\\ Johns Hopkins University, Baltimore, USA
}

\maketitle  
\begin{abstract}
\label{section:Abstract}
For minimally invasive endovascular surgery, the localization of catheters and guidewires inside the human body is essential. Electromagnetic (EM) tracking is one technology that allows localizing such surgical instruments. For localizing intra-operatively EM-tracked instruments with respect to preoperative volume data, it is necessary to bring pre- and intraoperative imaging into the same coordinate frame. In most existing solutions, such registration requires additional interactions, modifying the procedure's original workflow. We propose a new method taking advantage of Bioelectric signals to initialize and register preoperative volumes to the EM tracking system without significantly changing the interventional workflow. We envision the most natural use-case of our concept in cardiac electrophysiology (EP) procedures, in which EP catheters are already equipped with all the necessary sensing, including electric sensing for the measurement of electrophysiological signals and EM tracking for catheter localization. We use EP catheters for Bioelectric sensing to detect local features of the vasculature while advancing the catheter inside the human body. Such features can be automatically labeled before the procedure within the preoperative data. 
The combination of Bioelectric and EM tracking can localize vascular features such as bifurcations and stenosis within the EM tracking space. Mapping them to preoperative data automatically registers patients' CT space to EM tracking. The proposed registration process is entirely based on Bioelectric sensed features, with no need for external markers or other interventional imaging devices. 
In this paper, we showcase the proposed concept on a simplified vascular phantom and quantitatively analyze the initialization and registration results, applying the commonly used iterative closest point (ICP) path-based registration algorithms.

\end{abstract}

\section{Introduction}
\label{section:Introduction}
Treatment of cardiovascular diseases is often done in a minimally invasive endovascular approach. Such interventions require tubular instruments, like guidewires and catheters, to be advanced inside the human vascular system to the region of interest to apply the local treatment. In most cases, the surgeon uses one modality (sometimes multiple) for real-time visualization of the current catheter position, like fluoroscopy, ultrasound, or tracking devices such as electromagnetic (EM) catheter tracking \cite{abdelaziz_x-ray_2021}. In some cases, additional preoperative acquired volume data, e.g., the CT of the patient, is available. In order to leverage both pre- and intraoperative data during the procedure, the preoperative data has to be transformed into the same coordinate system as the intraoperative data. This transformation poses the problem of performing a (rigid) 3D registration.


\subsection{Related Work}
\label{section:RelatedWork}
\subsubsection{Electromagnetic Tracking}
Electromagnetic tracking is able to identify sensors in 5- or 6 DOF within a tracking field without the need of a line of sight. The location of these sensors is determined with respect to the field generator. There is no a priori method to translate this position to the vasculature of the patient's body, and the technology itself does not provide any imaging of anatomical features either \cite{franz_electromagnetic_2014}. Even though EM tracking is already implemented in commercial products and is a very well-researched technology in the scientific community, there still exists a hesitancy to utilize this technology interventionally \cite{franz_electromagnetic_2014}. Nonetheless, some successful implementations are listed following. One of the main problems of EM in this regard is the registration of the tracked catheter tip with the patient and the respective preoperative images. EM is especially difficult to implement in procedures that require registration in soft anatomical regions, such as the abdomen or thorax, and particularly in procedures that deal with the insertion of catheters into vasculature \cite{jackle_3d_2021,zhang_electromagnetic_2006}.

Different state-of-the-art research methods use different techniques to initialize and register the EM sensors to the subject and the respective preoperative data. First, the usage of external landmarks, such as fiducials or other markers attached to the subject when acquiring the preoperative image. These markers must be rigidly attached and not move between the preoperative image acquisition and the procedure. Such techniques have been presented by many authors, to perform initialization of registration \cite{jackle_3d_2021,lin_automatic_2020,wood_navigation_2005}. A second technique uses internal features or landmarks for registration, which can only be identified with interventional imaging modalities \cite{song_locally_2015}. One approach is to get a transformation by taking an image of the EM field generator and its position with respect to anatomical structures for registration  \cite{franz_electromagnetic_2014}.


\subsubsection{Bioelectric Navigation}
This concept is a recently introduced approach to navigation in minimally invasive endovascular procedures \cite{Sutton.2020}. Bioelectric navigation is based on the idea of detecting local geometric features of the vasculature. A catheter with a set of electrodes along its shaft is used, which generate a constant-amplitude AC current between a pair of these electrodes. The electric field associated with this current is influenced by the local geometry of the surrounding vasculature. The voltage between (the same or another) pair of electrodes along the shaft is measured, and the magnitude of this AC voltage at the stimulation frequency is extracted. As such, it produces a waveform containing features related to vascular geometry. Notably, changes in cross-sectional area (widening/narrowing of the vessel, e.g., from bifurcations, stenosis, and aneurysms) increase or decrease the measured voltages, thereby creating local minima and maxima in the extracted waveform. The authors proposed to use a preoperative segmented CTA model of the subject's vasculature and map the live waveform recorded by the catheter to expected reference waveforms obtained from the preoperative model, thus, identifying the catheter's position inside the vessel tree. Although Bioelectric Navigation can determine the current vascular branch and the catheter's proximity to geometric features, it does not provide a 3D location of the catheter. Also, it localizes the position only along the centerline, not detecting, e.g., a radial motion of the catheter. This potentially limits the applicability as a stand-alone modality.

\subsubsection{Electrophysiology-specific Registration}
As both EM tracking and the technical requirements for Bioelectric Sensing (multiple electrodes along the catheter) are already given for EP procedures, we see EP procedures as the most natural use-case of our framework. EP ablation procedures remain the gold standard for diagnosing and treating cardiovascular diseases associated with an irregular heart rhythm such as atrial fibrillation. During such procedures, EP catheters are commonly used to 1) create an anatomic 3D surface map of the atria and its electrical activity and 2) ablate the source regions causing irregular activity by applying radiofrequency energy, both by using the electrodes distributed over the catheter tip \cite{dong2006integrated}. A well known commercial mapping system, within this context, is the CARTO$^{\tiny \textregistered}$ system, which utilizes electromagnetic tracking that enables the computation of the 3D position of the catheter in the coordinate frame of the EM system and, as a result, enables the 3D surface reconstruction of the heart anatomy such as its chambers. The inclusion of preoperative data such as CT data as additional visualization and information can be further provided by computing a registration between CT and EM space and is incorporated into the CARTO$^{\tiny \textregistered}$ Merge module. For this purpose CARTO$^{\tiny \textregistered}$ offers two modalities, namely landmark-based registration and surface registration. The former utilizes manually labeled landmark pairs between the CT data and the activation map. In contrast, the latter uses the reconstructed surface map of the CARTO$^{\tiny \textregistered}$ system to register the surface to the corresponding surface extracted from the CT data \cite{bourier2014accuracy}. Compared to landmark-based registration, the procedure can be computed automatically. However, it also requires a map of the anatomy to be reconstructed interventionally prior to the actual registration. Similar, landmark-based registration procedures have been reported for ablation systems such as EnSite NavX \cite{richmond2008validation}.
In comparison to such systems, our method does not require an interventionally reconstructed surface of the anatomy for registration. However, it uses anatomical features that are automatically captured while traversing the catheter through the vasculature. After registration, the provided CT information can, in turn, already provide useful clinical information for improved navigation and activity map creation.

\section{Method}
\label{section:Method}
\subsubsection{Overview of the Method}
We propose a novel approach for initializing 3D EM tracking data registration with preoperative volume data using the sensing principle of Bioelectric Navigation. By this, we detect local geometrical vascular features along the catheter's path that are mapped to preoperative data and associated with the EM tracker's location. Based on these correspondences, we perform an initial registration which is later refined using ICP path-based registration.

\subsubsection{System Setup}
We used a diagnostic EP catheter (Boston Scientific EP XT) with $10$ electrodes located close to the tip.  For Bioelectric Navigation, the first and last electrodes (\#1 and \#10) were used as current injecting electrodes, and the inner pair (\#5 and \#6) as sensing electrodes. The waveform generator of a Picoscope 2203 digital oscilloscope was used to generate a sine wave voltage of $250 mV$ at $1kHz$. This voltage was transformed into an AC, constant amplitude current using a custom circuit and injected over the current injecting electrodes. The voltage between the sensing electrodes was digitized using the same oscilloscope. The sensed voltage, together with timestamps, is recorded using a python script and transformed into the Fourier domain using a sliding window FFT. Its absolute magnitude at the stimulation frequency was stored for further processing. A 5DOF EM Sensor (Aurora Flex Tube $1.0mm$, NDI) was attached to the catheter tip using heat shrink tubing at $2cm$ distance from the sensing electrodes. The EM tracking field is generated from an Aurora Field Generator Tabletop 50-70 (NDI). The EM tracking stream is recorded with timestamps along the path, using ImFusion Suite (ImFusion GmbH).

\subsubsection{Vascular Phantom} 
\begin{figure}[htp]
\centering
\subfloat[\label{fig:full_setup}]{%
  \includegraphics[width=0.68\textwidth]{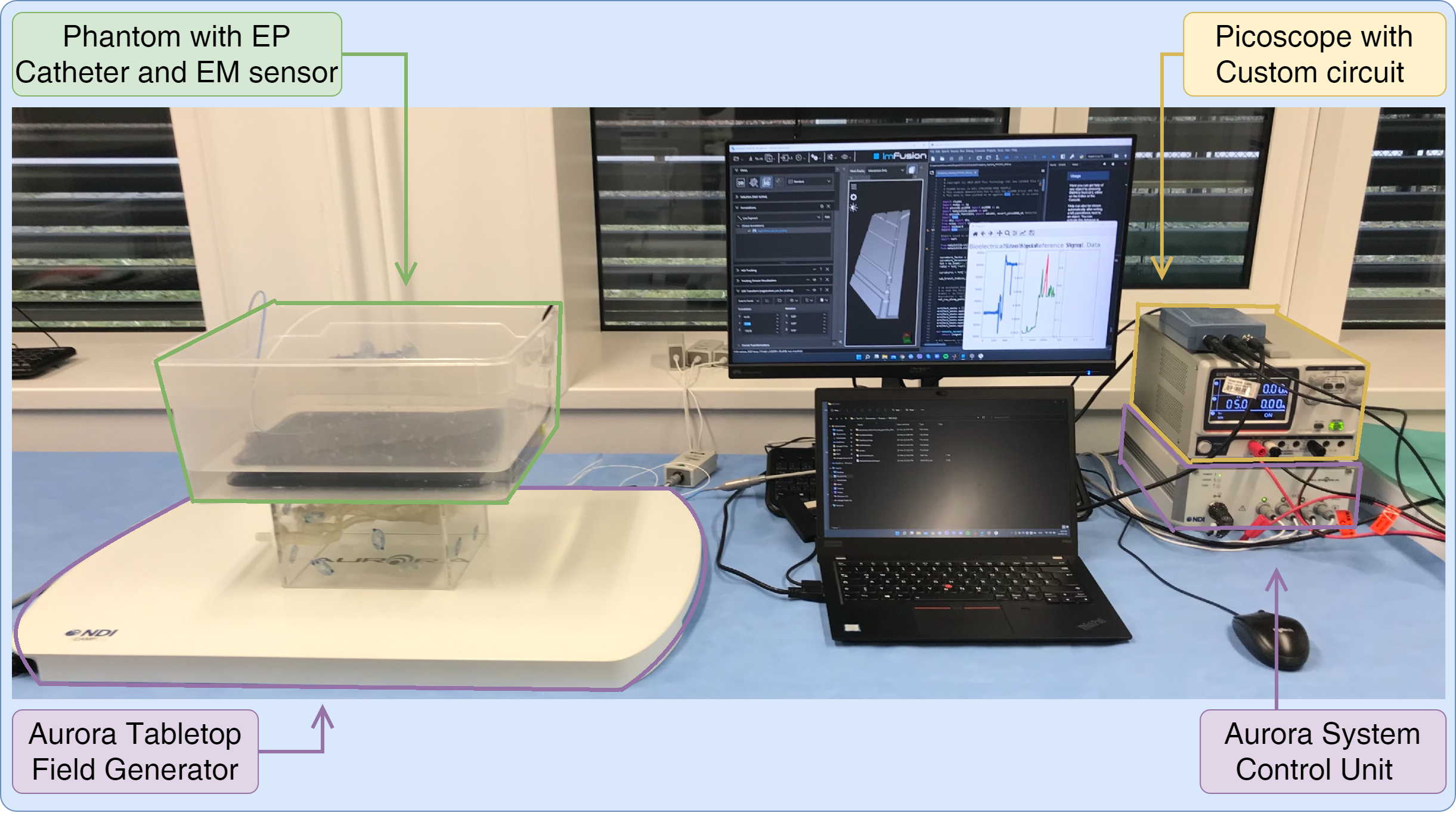}%
}\hfil
\subfloat[\label{fig:phantom_branches}]{%
  \includegraphics[width=0.27\textwidth]{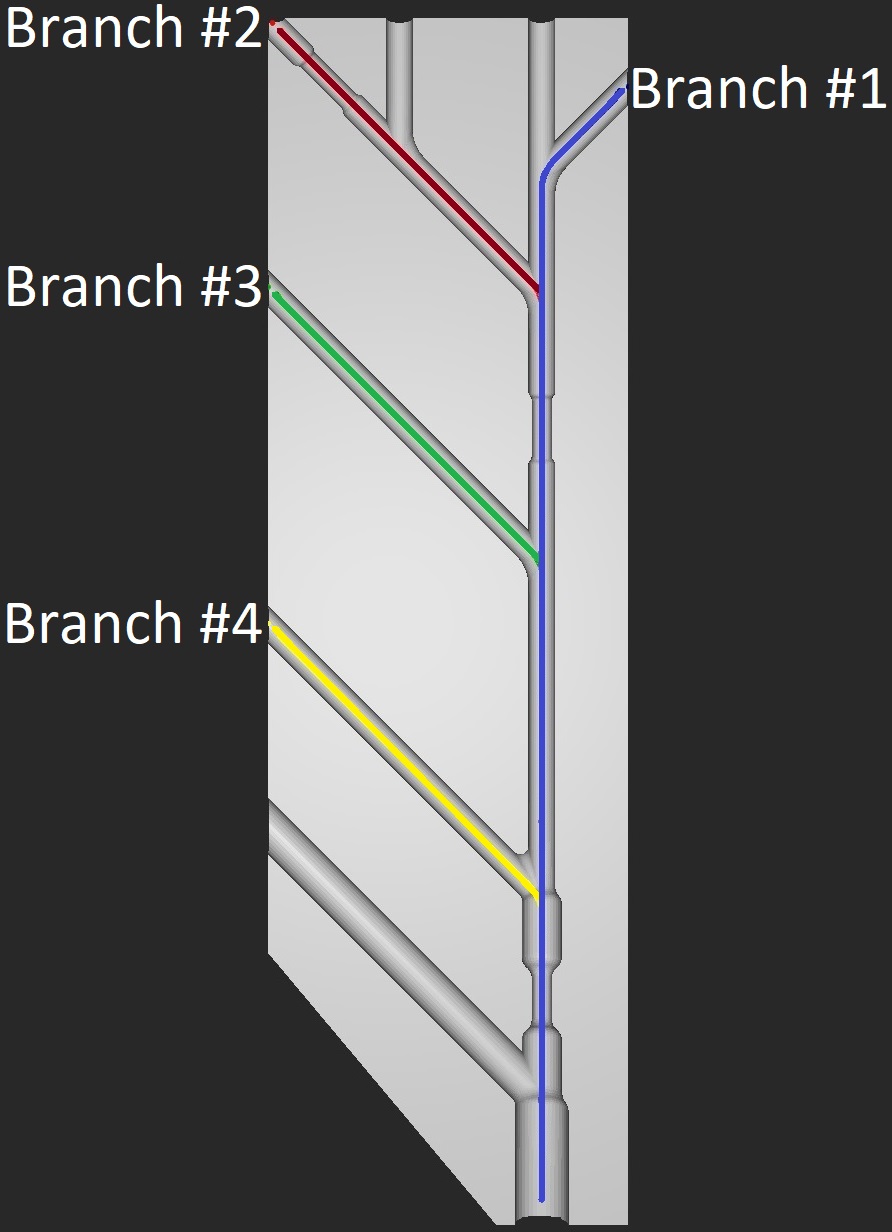}%
}

\caption{Full setup used for the experiments (a) and paths of the phantom traversed for the recordings (b)}
\label{fig:method_overview}

\end{figure}

For this prototypical study, an artificial, simplified vascular model was used, which we downloaded from a publicly available online repository \cite{Sutton.2020}. The STL model was cropped and upscaled to fit the catheter and EM tracker at once, and then the lower half was 3D printed in our lab. Although simplified, it contains relevant vascular features like different branches, bifurcations, and stenosis. The printed half of the phantom was then covered by a transparent plastic plate, leaving the vessel lumina as semicircles and allowing visual feedback on the catheter location. The phantom was submerged into a box containing a saline solution bath. The box was then fixed rigidly on top of the electromagnetic field generator. The complete setup can be seen in Fig.~\ref{fig:full_setup}, and a top view of the phantom's vascular structure is presented in Fig.~\ref{fig:phantom_branches}.

\subsubsection{Data Processing}
\begin{figure}[htp]
\centering
  \includegraphics[width=1.0\textwidth]{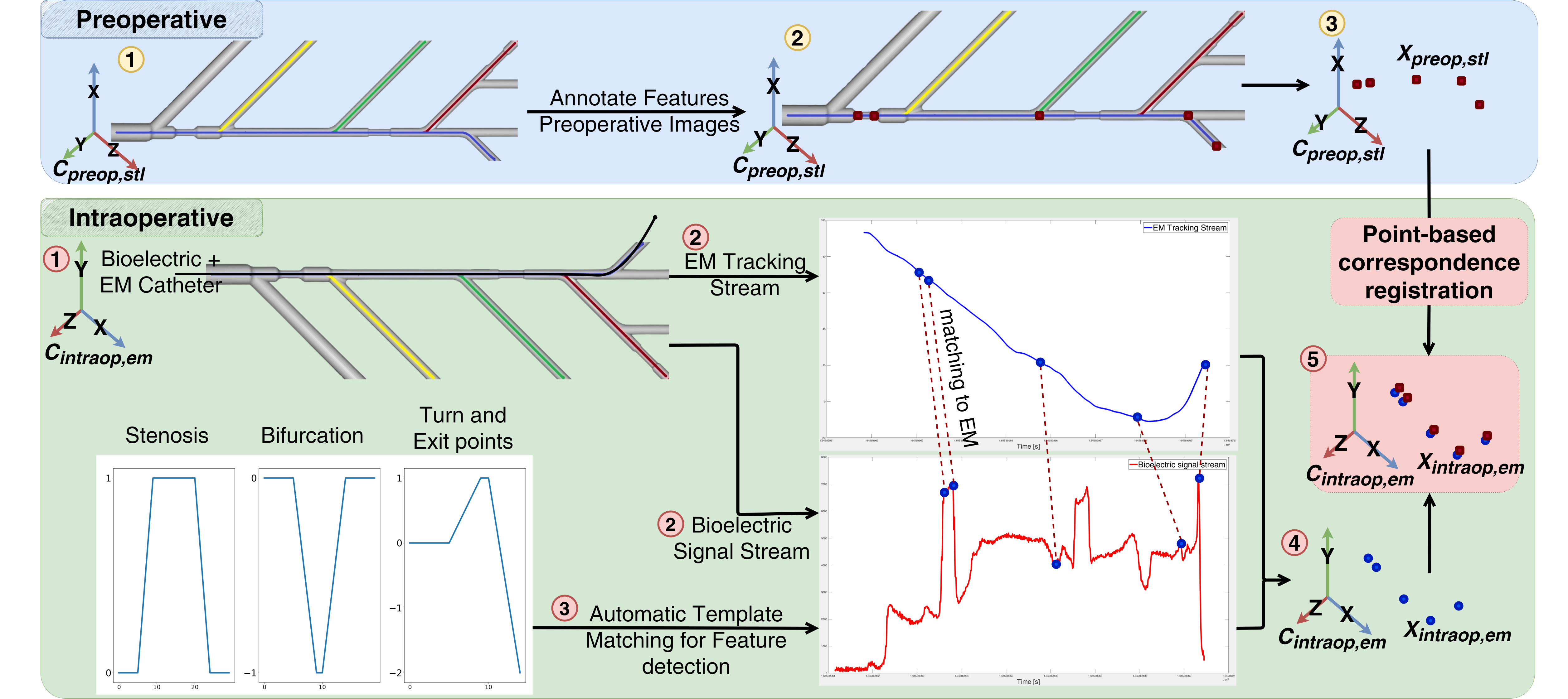}%
\caption{Overview of the registration pipeline. Before the intervention, preoperative centerline points $x_{pre,stl}$ are annotated on the pre-interventional model in local vascular landmarks. During the intervention, the features corresponding to these local vascular landmarks are annotated in the Bioelectric signal by template matching. Based on the timestamps of EM tracking data and Bioelectric signal, the corresponding intraoperative EM feature locations $x_{intra,em}$ can be extracted. Then, a rigid registration between these two point groups can be calculated using point-based correspondence.}
\label{fig:image2}
\end{figure}

The method is divided into preoperative and intraoperative steps. The workflow is summarized in Fig.~\ref{fig:method_overview}. We used the STL of the vascular phantom, representing a perfect reconstruction of the vascular geometry, in place of preoperative volume data. As preoperative steps, the centerlines of the four paths inside the phantom were extracted using 3D Slicer's SlicerVMTK toolkit (see Fig.~\ref{fig:phantom_branches}). These centerlines were visualized inside the STL using ImFusion Suite. We visually selected multiple points on the centerline corresponding to different vascular features (bifurcations/stenosis/turns/exit points) for each path. An exemplary annotation of such points can be seen in Fig.~\ref{fig:image2}. These points will from now on be referred to as preoperative centerline points $x_{preop,stl}$. The points are located in the preoperative coordinate frame $C_{preop,stl}$. The surgeon must label at least three features in the preoperative CT as a preoperative step so that the Bioelectric sensing and EM can match and register these features during the intervention.

The first intraoperative step is the insertion of the catheter in the phantom together with the attached EM tracker. We recorded the data streams while pulling the catheter from the beginning of the vascular tree towards the endpoints. After traversing a path, the Bioelectric sensing and EM tracking streams were stored together with their timestamps. Then, the EM tracking stream was filtered with a Gaussian filter (window width of $25$ samples; filtering in both rotation and translation) to account for noise in the tracking signal. The Bioelectric signal was filtered with a 5 sample uniform filter for smoothing. Next, for each of the preoperative centerline points $x_{preop,stl}$, their corresponding Bioelectric feature was located by a template matching algorithm. The template matching uses the annotations from the preoperative step, namely the landmark locations in the STL as well as the type of landmark (bifurcation/stenosis/turn into branching vessel/exiting point of phantom). The templates have been defined as simplistic templates based on estimating the vessel's cross-sectional area, as calculated from the maximum inscribed sphere radii extracted by VMTK. The templates are depicted in Fig.~\ref{fig:image2}. The algorithm first identifies one salient feature, i.e., which is easy to recognize. In our implementation, we chose the first stenosis that is present in all paths. We extract sliding windows, normalize them by their mean and standard deviation and search the location of maximum correlation with a stenosis template within the first 12 cm of the Bioelectric signal. Based on the annotations in the preoperative data, we calculate the distance of all other vascular landmarks to this salient feature. Then we extract small regions of the signal (calculated landmark location $\pm$ 1.5 cm) and perform template matching there, selecting the locations with the highest correlation as the feature location within each given region. From these locations in the Bioelectric signal, the timestamp is extracted and used to find corresponding EM tracking points. As the Bioelectric and EM streams were recorded with different sampling frequencies, the closest timestamp in the EM stream was chosen as the associated timestamp for the given Bioelectrical timestamp. For each of these timestamps, the corresponding EM location was selected. We denote this intraoperative location as $x_{intraop,em-offset}$. In order to account for $2.0cm$ offset between the EM sensor and Bioelectric sensing electrodes, starting from this location $x_{intraop,em-offset}$, the EM tracking stream was traversed in a negative time direction, adding up the Euclidean distances of sequential locations until reaching a sum of $2.0 cm$. This resulting offset-corrected EM location was chosen as the location of the EM tracker $x_{intraop,em}$, located in the intraoperative EM coordinate frame $C_{intraop, em}$.
 
These steps render two sets of points, $x_{preop,stl}$ and $x_{intraop,em}$, for which a point-to-point correspondence is known through the Bioelectrical feature matches. A point-correspondence-based registration is performed in ImFusion Suite to obtain an initial registration. 
We further refined this initial registration using an ICP path-based registration between the initially registered EM tracking stream and the extracted centerlines of the chosen path \cite{Pomerleau.2013,nypan_vessel-based_2019}. As the path-based registration matches all points of the EM tracking stream to centerline points, this step was performed to readjust the registration for any uncertainty in the feature detection from the Bioelectric stream. The ICP registration produced the final registration used to register the preoperative space to the intraoperative EM tracking space.

\section{Evaluation}
\label{section:Evaluation}
\subsection{RMSE of registrations with centerline}
We quantitatively evaluate the Root Mean Square Error (RMSE) error of the registration. For this, recordings of EM tracking and the Bioelectric signal from $4$ branches were taken, with three runs for each branch. For each of these $12$ recordings, we evaluated the RMSE in two different scenarios. Firstly, we calculated the RMSE after initial registration between centerline labeled features $x_{pre,stl}$ and Bioelectric features (respective EM positions) $x_{intra,em}$. Secondly, we calculated the RMSE after full registration (initialized as above and refined by ICP) between the EM recorded stream and respective centerline. The per-path averages and standard deviations of these RMS errors are depicted in Table \ref{Tab:RMSE}, columns ``Initialization vs. Centerline" and ``ICP-Refined vs. Centerline".

\subsection{RMSE of registration results and ground truth} 
We also evaluated the RMSE as compared to a ground truth registration. We extracted positions at the corners of the actual phantom for the ground truth registration by visually touching them with the EM tracker. Then, a point-based transformation was performed between EM acquired corner locations and the respective STL corners. We calculate the RMSE between the original EM path against the ground truth transformation of the previous approaches (Initialization and ICP-Refined resulted registrations). The results of these comparisons can be seen in Table~\ref{Tab:RMSE}, columns ``Initialization vs. Ground Truth" and ``ICP-Refined vs. Ground Truth''.

\begin{table}[]
\caption {Mean and standard deviation of RMSE of the two registration steps for each branch, as well as compared to the ground truth registration. Results given in millimeters.}
\center
\begin{tabular}{c|cc|cc|cccccccc}
\hline\hline
\multirow{3}{*}{Branch \#} &
  \multicolumn{2}{c|}{\begin{tabular}[c]{@{}c@{}}Initialization\\ vs. Centerline\end{tabular}} &
  \multicolumn{2}{c|}{\begin{tabular}[c]{@{}c@{}}ICP-Refined\\ vs. Centerline\end{tabular}} &
  \multicolumn{4}{c|}{\begin{tabular}[c]{@{}c@{}}Initialization\\ vs. Ground Truth\end{tabular}} &
  \multicolumn{4}{c}{\begin{tabular}[c]{@{}c@{}}ICP-Refined\\ vs. Ground Truth\end{tabular}} \\
 &
  \multicolumn{2}{c|}{Points} &
  \multicolumn{2}{c|}{Paths} &
  \multicolumn{2}{c|}{Points} &
  \multicolumn{2}{c|}{Paths} &
  \multicolumn{2}{c|}{Points} &
  \multicolumn{2}{c}{Paths} \\ 
 &
  \multicolumn{1}{c|}{mean} & 
  \multicolumn{1}{c|}{$\sigma$} &
  \multicolumn{1}{c|}{mean} &
  \multicolumn{1}{c|}{$\sigma$} &
  \multicolumn{1}{c|}{mean} & 
  \multicolumn{1}{c|}{$\sigma$} &
  \multicolumn{1}{c|}{mean} &
  \multicolumn{1}{c|}{$\sigma$} &
  \multicolumn{1}{c|}{mean} & 
  \multicolumn{1}{c|}{$\sigma$} &
  \multicolumn{1}{c|}{mean} &
  \multicolumn{1}{c}{$\sigma$}\\ \hline\hline
Branch 1 &
  \multicolumn{1}{c|}{2.77} &
  \multicolumn{1}{c|}{0.07} &
  \multicolumn{1}{c|}{0.97} &
  \multicolumn{1}{c|}{0.04} &
  \multicolumn{1}{c|}{1.72} &
  \multicolumn{1}{c|}{0.05} &
  \multicolumn{1}{c|}{1.74} &
  \multicolumn{1}{c|}{0.10} &
  \multicolumn{1}{c|}{2.72} &
  \multicolumn{1}{c|}{0.49} &
  \multicolumn{1}{c|}{2.74} &
  \multicolumn{1}{c}{0.46} \\ \hline
Branch 2 &
  \multicolumn{1}{c|}{2.06} &
  \multicolumn{1}{c|}{1.09} &
  \multicolumn{1}{c|}{1.04} &
  \multicolumn{1}{c|}{0.07} &
  \multicolumn{1}{c|}{2.50} &
  \multicolumn{1}{c|}{0.71} &
  \multicolumn{1}{c|}{2.66} &
  \multicolumn{1}{c|}{0.99} &
  \multicolumn{1}{c|}{2.21} &
  \multicolumn{1}{c|}{0.15} &
  \multicolumn{1}{c|}{2.26} &
  \multicolumn{1}{c}{0.11} \\ \hline
Branch 3 &
  \multicolumn{1}{c|}{1.70} &
  \multicolumn{1}{c|}{0.31} &
  \multicolumn{1}{c|}{0.82} &
  \multicolumn{1}{c|}{0.05} &
  \multicolumn{1}{c|}{2.64} &
  \multicolumn{1}{c|}{0.32} &
  \multicolumn{1}{c|}{2.62} &
  \multicolumn{1}{c|}{0.29} &
  \multicolumn{1}{c|}{3.17} &
  \multicolumn{1}{c|}{0.05} &
  \multicolumn{1}{c|}{3.20} &
  \multicolumn{1}{c}{0.05} \\ \hline
Branch 4 &
  \multicolumn{1}{c|}{1.71} &
  \multicolumn{1}{c|}{0.25} &
  \multicolumn{1}{c|}{0.78} &
  \multicolumn{1}{c|}{0.07} &
  \multicolumn{1}{c|}{2.44} &
  \multicolumn{1}{c|}{0.14} &
  \multicolumn{1}{c|}{2.54} &
  \multicolumn{1}{c|}{0.12} &
  \multicolumn{1}{c|}{2.21} &
  \multicolumn{1}{c|}{0.20} &
  \multicolumn{1}{c|}{2.10} &
  \multicolumn{1}{c}{0.18} \\ \hline\hline
\textbf{Overall} &
  \multicolumn{1}{c|}{\textbf{2.06}} &
  \multicolumn{1}{c|}{\textbf{0.50}} &
  \multicolumn{1}{c|}{\textbf{0.90}} &
  \multicolumn{1}{c|}{\textbf{0.12}} &
  \multicolumn{1}{c|}{\textbf{2.33}} &
  \multicolumn{1}{c|}{\textbf{0.41}} &
  \multicolumn{1}{c|}{\textbf{2.39}} &
  \multicolumn{1}{c|}{\textbf{0.44}} &
  \multicolumn{1}{c|}{\textbf{2.58}} &
  \multicolumn{1}{c|}{\textbf{0.46}} &
  \multicolumn{1}{c|}{\textbf{2.57}} &
  \multicolumn{1}{c}{\textbf{0.50}} \\ \hline\hline
\end{tabular}
 \label{Tab:RMSE}
\end{table}

\section{Discussion}
\label{section:Discussion}
When evaluating the initial registration and the final, path-based registration against the centerlines, it can be seen that the initial registration already achieves satisfactory results with an RMSE of $2.06mm$ throughout all branches. The path-based ICP further reduces this error to $0.9mm$, showing the utility of this refinement technique. Compared to this, RMSE calculation against the ground truth registration shows overall increased RMSE errors. Also, it can be seen that the relative benefit of the ICP refinement actually leads to a slight decrease in registration performance under this metric. We hypothesize that the reason is mostly that the EM tracker was attached to the outside of the catheter, thus, being radially offset from the catheter's center by design. As the Bioelectric initialization assumes that the visited points are on the centerline and the ICP registers to the centerline, they tend to force their registration into the vessel's center. At the same time, the radial offset of the tracker leads to it being closer to a vessel wall. We expect that this error could be minimized by registering to paths that minimize the distance to the vessel's wall and not the centerline \cite{de_lambert_electromagnetic_2012}.

As for the template matching, the current implementation assumes that the path is already traversed (i.e., is known). If this cannot be guaranteed, the path could be classified firstly, using the Bioelectric signal as in \cite{Sutton.2020}. 

As discussed, most current solutions propose methods to initialize the registration process by either changing the interventional workflow, utilizing additional external imaging modalities, or registration markers. The method we propose utilizes the same technologies as for electrophysiology (EP) procedures, with minimal modifications, like a signal generator for the Bioelectric sensing. Furthermore, the method does not affect the intraoperative workflow, providing an initialization of the registration after traversing with the Bioelectric sensing catheter via a couple of vascular landmarks.
\section{Conclusions}
\label{section:Conclusion}
In this paper, we proposed a pipeline for registration of EM tracking space to preoperative data using features extracted from Bioelectric Navigation alone, as well as together with a path-based registration. The approach could successfully register EM tracking space and preoperative data in a simplified vascular phantom. Questions that arise for future research are how well this approach can perform in more realistic vascular phantoms like the anatomy found close to the heart, which would be relevant for EP procedures. Also, an evaluation of the performance in real tissue will be of interest. We see a natural path for our concept in the application to EP procedures, where EM tracking technology is already in use and catheters are equipped with the electrodes necessary for Bioelectric sensing.

\section*{Acknowledgments}
\label{acknowledgments}
This project has been funded from the grant \textit{``Digitaler OP''} by the Bayerisches Staatsministerium f\"{u}r Wissenschaft und Kunst (1530/891 02). H. Maier was supported by TUM International Graduate School of Science and Engineering (IGSSE) and the ICL-TUM Joint Academy of Doctoral Studies (JADS) program. N. Navab was partially supported by U.S. National Institutes of Health under grant number 1R01EB025883-01A1.

%
%
%
\bibliographystyle{splncs04}
\bibliography{bibliography}
\end{document}